\documentclass[conference]{IEEEtran}
\IEEEoverridecommandlockouts

\usepackage{ps}

\preprinttrue
\publishedfalse

\renewcommand{\copyrightOwnerFull}{European Control Association (EUCA)}
\renewcommand{\copyrightOwnerShort}{EUCA}

\input{copyright.tex}

\usepackage{cite}
\usepackage{amsmath,amsthm, amssymb, amsfonts}
\usepackage{array}
\usepackage{multicol}
\usepackage{orcidlink}
\usepackage{algorithm}
\usepackage{algpseudocode}
\usepackage{graphicx}
\usepackage{textcomp}
\usepackage{comment} 
\usepackage{xcolor}
\def\BibTeX{{\rm B\kern-.05em{\sc i\kern-.025em b}\kern-.08em
    T\kern-.1667em\lower.7ex\hbox{E}\kern-.125emX}}
\usepackage{glossaries}

\newglossaryentry{realnumbers}{
  name={\ensuremath{\mathbb{R}}},
  description={the set of real numbers},
  sort={real numbers}
}

\newglossaryentry{RWTH}{
  type=\acronymtype,
  name={RWTH},
  description={Rheinisch-Westfälische Technische Hochschule Aachen},
  first={Rheinisch-Westfälische Technische Hochschule Aachen (RWTH)},
  sort={Rheinisch-Westfälische Technische Hochschule Aachen}
}
\newacronym{SOMC}{SOMC}{Service-Oriented Model-Based Control}
\newacronym{SOA}{SOA}{Service Oriented Architecture}
\newacronym{ASOA}{ASOA}{Automotive Service-Oriented Architecture}
\newacronym{TOPSIS}{TOPSIS}{Technique for Order Preference by Similarity to Ideal Solution}
\newacronym{QoS}{QoS}{Quality of Service}
\newacronym{SOC}{SOC}{Service Oriented Computing}
\newacronym{MPC}{MPC}{Model Predictive Controller}
\newacronym{AI}{AI}{Artificial Intelligence}
\newacronym{IoT}{IoT}{Internet of Things }
\usepackage{tabularx}
\usepackage{import}
\usepackage{newtxtext,newtxmath}
\usepackage{multirow}
\theoremstyle{definition}
\newtheorem{definition}{Definition}
\newtheorem*{remark}{Remark}
\usepackage{csquotes}
\usepackage{mathtools}
\usepackage{bm} 

\begin{document}
\title{Graph-Based Orchestration of Service-Oriented Model-Based Control Systems
{\footnotesize \textsuperscript{}}
\thanks{This research is supported by the Deutsche Forschungsgemeinschaft
(German Research Foundation) with the grant number 468483200.}
\thanks{
            $^{1}$ The authors are with the Department of Aerospace Engineering, University of the Bundeswehr Munich, Germany, {\texttt{\{firstname\}.\{lastname\}@unibw.de}}}
\thanks{
            $^{2}$ The author is with the Institute of Automatic Control, RWTH Aachen University, Germany, {\texttt{l.doerschel@irt.rwth-aachen.de}}}}

\author{Julius Beerwerth$^{1}$\,\orcidlink{0000-0001-6167-9692}, Hazem Ibrahim$^{1}$\,\orcidlink{0009-0007-1911-4760},~\IEEEmembership{Student~Members,~IEEE}, Bianca Atodiresei$^{1}$,\\
Lorenz Dörschel$^{2}$\,\orcidlink{0000-0001-9566-6371}, ~\IEEEmembership{Member,~IEEE}, and Bassam Alrifaee$^{1}$\,\orcidlink{0000-0002-5982-021X},~\IEEEmembership{Senior Member,~IEEE}
}

\maketitle
\begin{abstract}
This paper presents a novel graph-based method for adapting control system architectures at runtime. We use a service-oriented architecture as a basis for its formulation. In our method, adaptation is achieved by selecting the most suitable elements, such as filters and controllers, for a control system architecture to improve control systems objective based on a predefined cost function. Traditional configuration methods, such as state machines, lack flexibility and depend on a predefined control system architecture during runtime. Our graph-based method allows for dynamic changes in the control system architecture, as well as a change in its objective depending on the given system state. Our approach uses a weighted, directed graph to model the control system elements and their interaction. In a case-study with a three-tank system, we show that by using our graph-based method for architecture adaptation, the control system is more flexible, has lower computation time, and higher accuracy than traditional configuration methods.
\end{abstract}


\section{Introduction}
\label{sect:intro}
\subsection{Motivation}
\glspl{SOA} partition software into smaller, independent units called services, which can be distributed across different hardware systems \cite{b18}. Services follow common principles like reusability and composability, allowing them to form service compositions\cite{b18}. The process of creating and adapting service compositions is called orchestration, and it is managed by a central entity known as the orchestrator. Orchestration plays a central role in \gls{SOA} by managing and connecting services to have functions tailored to system needs\cite{b18}.

The application of \gls{SOA} concept to control systems architectures is different from the typical use of \gls{SOA} in web technology for example. This difference results from the real-time requirements of control systems. This application forms an adaptable control system architecture at run-time, so any element of the control system can be changed. This advantage of adaptability can result in having a better control system's response compared to a control systems with fixed architectures. We apply the concept of \gls{SOA} to control system  architecture by modeling the  elements of a control system as services and integrate them at run-time using the orchestrator. In our previous work, we presented an example for the implementation of this concept which is a \gls{SOMC} Architecture \cite{b13}. Our implementation is based on the \gls{ASOA} presented by Kampmann et al.\cite{kampmann2022asoa}. An advantage of \gls{ASOA} is that it supports updatability of the embedded software\cite{b41}. This paper extends this concept by providing a novel orchestration method. 

In \cite{b13}, the orchestrator has system-level knowledge of all the control system services. The orchestrator adapts the architecture based on a user-defined heuristic, for example favoring the newest services. In this case, when a new service is detected, it replaces the previous service of the same type. However, this heuristic does not consider the performance of the control system architecture, as newer services may not always deliver the best control system performance. While offering more flexibility than traditional systems such as state machines, the previous used orchestrator does not account for potential changes in the control system objective which is a predefined cost function in our work.

We address this issue with our graph-based orchestration approach by evaluating which service should be used in the control system architecture, based on the control system's objective. The control system's objective is predisposed to change, and the orchestrator can handle it accordingly. Moreover, the orchestrator's ability to react to changes in objective significantly increases the adaptability of the control system\cite{b13}. We also distinguish between function and resource orchestration \cite{b40}. In this paper we focus on function orchestration.  

\subsection{Related Work}

In control theory, several methods aim to create runtime-flexible control systems. Switching adaptive control can detect changes and toggle between pre-defined controllers \cite{b28}. For overactuated systems, control allocation redistributes control values to compensate for actuator failures \cite{b29}. However, these techniques predetermine all components and their interactions during the design phase of the control system, limiting flexibility to a set of predefined options.
An example of adapting control system architectures is the dynamic updating of control systems for evolving self-adaptive systems introduced in \cite{b30}. This approach focuses on identifying changes in the objective and then constructing control systems using components that offer dynamic control or deployment methods. Another approach is the Metacontrol for the Robot Operating System (MROS) framework. It is a model-based system for real-time adaptation of robot control system architectures using ROS. It combines domain-specific languages to model various architectural options. MROS also implements the MAPE-K cycle (Monitor, Analyze, Plan, Execute, Knowledge) and meta-control frameworks for dynamic adaptation using an ontology-based approach \cite{b31}.

In the domain of \gls{SOA}, multiple approaches have been proposed to use graph-based methods to solve optimal service composition. For example, \cite{b10} implemented a graph-based orchestration method using web services and an \gls{AI} graph-planning algorithm to select optimal service paths. \cite{b11} employed k-dimensional trees and nearest neighbor search for cloud service selection, demonstrating fast, and effective service composition. H. Alhosaini et al. \cite{b20} proposed a hierarchical method based on skyline services. Skyline services are a type of service that acts as a gateway or intermediary between client applications and other backend services. This method precomputes Pareto-optimal skylines to optimize \gls{QoS}-driven service composition, improving efficiency through precomputation and caching. In Mobile Edge Computing, J. Wu et al. \cite{b15} introduced M3C, an optimization method for micro-service composition that minimizes latency and energy use, offering practical deployment benefits.
Beside this method, \cite{b23} developed a top-k QoS-optimal service composition method for \gls{IoT} systems, leveraging service dependency graphs and dynamic programming to reduce search complexity and improve performance. Similarly, \cite{b24} proposed a heuristic polynomial-time graph search algorithm for Web Service Composition. \cite{b25} introduced the Pre-joined Service Network, which uses graph databases to retrieve optimal service compositions efficiently.
Control system architectures traditionally rely on fixed component interactions \cite{b19}. Approaches like switching adaptive control \cite{b28} offer some runtime flexibility but remain limited. \gls{SOA} opens new possibilities for more dynamic and flexible control system architectures. The previous discussed methods focus on the usage of \gls{SOA} mostly in web technologies which does not include real-time requirements. Our paper fills this gap by using \gls{SOA} in control systems which adds the real-time challenge. This challenge is addressed by modeling the control system elements, such as filters and controllers, as services and adapting them at runtime using our graph-based approach.

\subsection{Paper Structure}
The remainder of this paper is structured as follows: Section \ref{sect:methimpl} discusses our graph-based approach consisting of two phases: service graph creation and graph-based orchestration, Section \ref{sect:evaluation} elaborates on the evaluation scenario using a three tank control system. Section \ref{sect:conclusion} provides an outlook on future work.

\section{Methodology and Implementation}
\label{sect:methimpl}
This section provides a step-by-step explanation of the concept of graph-based orchestration on a control system using \gls{SOA}. Section \ref{sect:graphcreation} presents the service graph, the data structure used as the foundation of our orchestration method. Section \ref{sect:orchworkflow} presents an explanation on how the orchestrator uses this service graph to establish control system architectures.

\subsection{Service Graph Definition}
\label{sect:graphcreation}
In computer science, graphs are often used to model communication networks \cite{b8}. A graph is composed of a set of nodes and a set of edges that connect the nodes. Depending on the nature of the edges, a graph is directed or undirected. Further, a graph is either weighted or unweighted. A sequence of nodes connected by (weighted) edges form a (weighted) path. The shortest path is the one with the smallest sum over the weights \cite{b9}. Dijkstra's algorithm addresses the problem of finding the shortest path in a weighted graph\cite{b16}.

\begin{figure}
    \centering
    \includegraphics[width=0.45\textwidth]{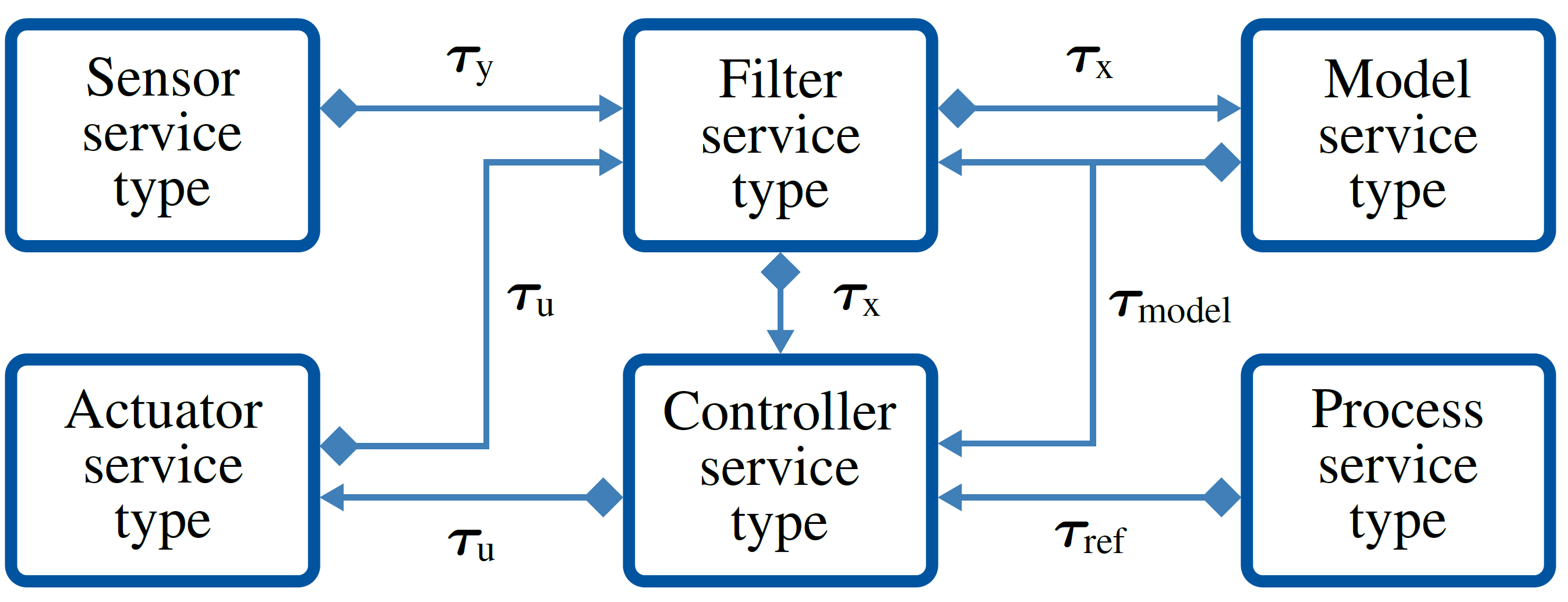}
    \caption{Control system respresentation as graph of services.}
    \label{fig:basic_control_loop}
\end{figure}

In the context of \gls{SOA}, graphs are good representations  for service composition and orchestration forming what is called service graph. Nodes represent services, while edges indicate communication between them. 
Figure~\ref{fig:basic_control_loop} presents a generalized architecture of the control system represented as a service graph. The control system architecture consists of sensor, actuator, filter, controller, model, and process. Here, the process service is not an abstraction of the real process, but simply provides the state references. The interfaces to the real process are the sensor and the actuator services, closing the control loop. Each control system element is modelled as a service to facilitate the use of \gls{SOA}. In our \gls{SOMC} architecture, a service is defined by its interfaces and tasks \cite{kampmann2022asoa}. Interfaces enable the exchange of data between services and tasks enable the implementation of functions. The interfaces of a service are partitioned into requirements and guarantees. A requirement may be interpreted as an input representing information, that is required by the service. A guarantee may be interpreted as an output, representing information that is provided by the service. The type of data provided and required by a guarantee and a requirement, respectively, is encoded in what is called functionality type. A requirement and a guarantee are compatible if they have the same functionality type.
 For example, the filter service requires measured values of the functionality types  $\tau_y$, $\tau_{u}$, and a model of the functionality type $\tau_{model}$ and guarantees to provide the system state of the functionality type $\tau_x$.
 
The control system architecture in Figure~\ref{fig:basic_control_loop} forms a directed service graph. Each block represents a service and can be attributed to a node in the service graph. The relationships between the services is depicted by directed edges in the service graph.

The goal of graph-based orchestration is to choose the optimal control system architecture at any given time. An optimal control system architecture is defined as the service composition that has the lowest cost based on a predefined cost function. In our case, this is represented by the shortest path from the sensor to the actuator.
\begin{definition}[Control System]
The control system is defined by sensors, filters, controllers, and actuators forming a control loop.
\end{definition}
\begin{definition}[Control System Objective]
The control system objective is encoded in a predefined cost function. 
\end{definition}
\begin{definition}[Optimal Control System Architecture]
    We define optimality in \gls{SOA} as the shortest path in the service graph based on the cost function.
\end{definition}
To apply Dijkstra's algorithm to the service graph in Figure \ref{fig:basic_control_loop}, some adjustments must be made in order to transform this graph.
\begin{enumerate} 
    \item We add a fixed start node and fixed target node. The start node is connected to all available sensor nodes. All available actuator nodes are connected to the target node.\label{l}
    \item We remove the process service as we assume it is fixed and cannot be changed by the orchestrator. Note, that the process service only provides the state reference and is not an abstraction for the real process. The control loop remains closed.
    \item Controllers and filters may rely on a model. If a model is used, it must be considered in the selection of the optimal controller or filter. As depicted in Figure~\ref{fig:basic_control_loop}, the model is not always part of the direct path from the sensor to the actuator. In some cases, the shortest path bypasses the model, linking the filter directly to the controller. Alternatively, a cycle may form between the filter, models, and controller, creating ambiguity regarding which model the controller should use. To resolve this, filters/controllers and the respective model are grouped into a single node.
    \item We remove the edge from the actuator to the filter as it introduces a cycle into the service graph. While Dijkstra's algorithm is able to handle cycles, it is not able to handle negative weights. So assuming non-negative weights we can omit all edges that introduce a cycle as they cannot be part of the shortest path.
\end{enumerate}

\begin{figure*}
  \centering
  \includegraphics[width=0.80\textwidth]{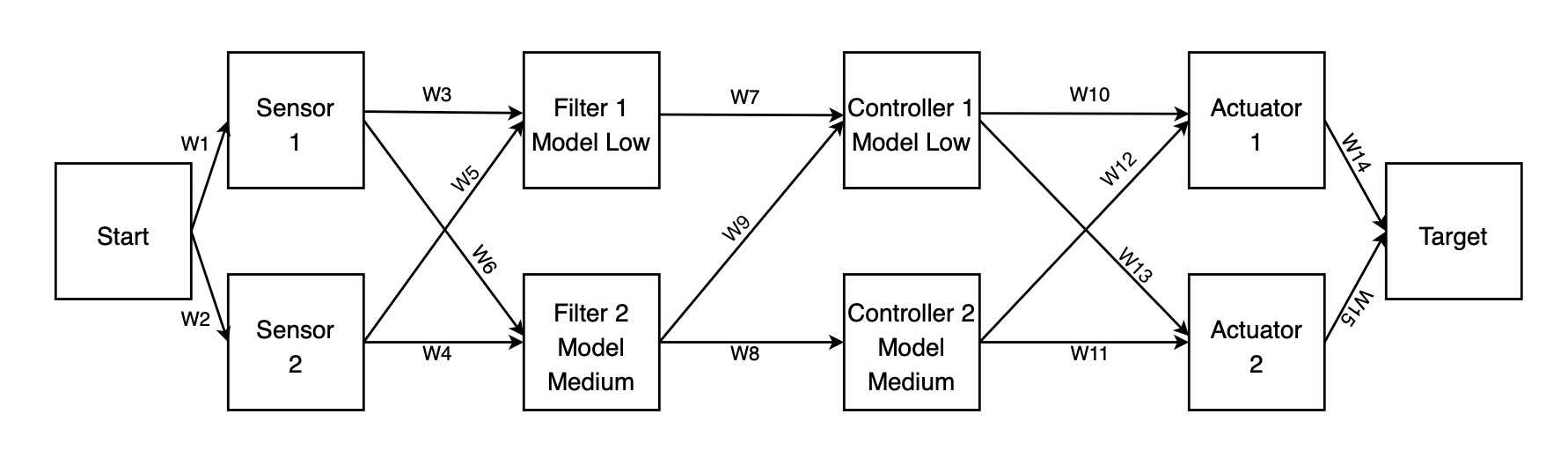}
  \caption{Example of control system architecture graph with start and target nodes, and instantiated service types. Edges show varying model complexities.}
  \label{fig:starttarget}
\end{figure*}

Applying these steps to the service graph example in Figure \ref{fig:basic_control_loop} and assuming two available services for each control system element results in the service graph in Figure \ref{fig:starttarget}. 

In this paper, we assume linear and time-invariant models. For this scope, we define three types of possible models: a low complexity model, a medium complexity model, and a high complexity model. The complexity of the model is determined by the size of its system matrix $\mathbf{A}$.
\begin{definition}[Model Complexity]
     Linear and time-invariant (LTI) models are representable as matrices in the relation \begin{align}
     \dot{\mathbf{x}} &= \mathbf{A} \mathbf{x} +\mathbf{B} \mathbf{u}
     \end{align}where $\mathbf{x}\in  \mathbb{R}^n$ and $\mathbf{u}\in  \mathbb{R}^m$ represent the system states and inputs, respectively and $\mathbf{A}\in \mathbb{R}^{n\times n}$ and $\mathbf{B}\in \mathbb{R}^{n\times m}$ are the system matrix and the input matrix. The model's complexity is proportional to the dimension of the system matrix $\mathbf{A}$. 
\end{definition}
Here, we categorize model complexity into three distinct levels: low, medium, and high.
Note that in the example in Figure \ref{fig:starttarget} the edge from Filter 1 using the low model to Controller 2 using the medium model is one of the missing edges.
In general, filter and controller services may be incompatible due to different model structures. The structure of the model influences the potential connection between the filters and controllers that use these specific models. A controller using a higher complexity model cannot function properly with input from a lower complexity filter. The reason behind this is that a controller requires a system state vector of higher complexity, e.g., $\mathbf{x_{\text{controller}}} \in \mathbb{R}^5$ containing five states cannot work with the output of a filter that provides a system state vector of lower complexity, e.g., $\mathbf{x_{\text{controller}}} \in \mathbb{R}^3$ containing only three states. Additionally, a filter with a high complexity model is only compatible with a controller with a low complexity model, if the low complexity state is a subset of the high complexity state. In this work we assume that every higher complexity state includes a subset of all states with lower complexities. Therefore, we simplify the compatibility problem between filter and controller to the following constraint applied in Figure \ref{fig:starttarget}. An edge between a filter and a controller is possible if and only if the complexity of the filter model is higher than or equal to the complexity of the controller model.

The graph-based orchestrator aims to select the optimal combination of sensors, filters, controllers, and actuators. To enable this, each edge in the service graph is assigned a weight, allowing for the application of shortest path algorithms. The edge weight represents the cost of choosing the service it points to. These weights are determined by the cost function.

We use the following cost function:
\begin{align}
    \text{cost} &= (\alpha_{\text{comp}} \ \beta_{\text{inacc}}) \begin{pmatrix} x_\text{comp} \\ y_\text{inacc} \end{pmatrix}\text{,}
    \label{eq:cost_function}
\end{align}
where:
\begin{itemize}
    \item the \textit{computation factor} $x_\text{comp}$$\in \mathbb{R}_{>0}$ and the  \textit{inaccuracy} $y_\text{inacc}$ $\in \mathbb{R}_{>0}$ depend on the service. The \textit{computation factor} reflects the computation time. Lower values for both $x_\text{comp}$ and $y_\text{inacc}$ are preferable, as they signify lower costs, faster task completion, and higher precision.
    \item $\alpha_\text{comp}$ and $\beta_\text{inacc}$ $\in \mathbb{R}_{>0}$ are weighting factors that determine the relative importance of the variables $x_\text{comp}$ and $y_\text{inacc}$ in the total cost computation. The higher $\alpha_\text{comp}$ and $\beta_\text{inacc}$, the more importance is placed on the respective variable. 
\end{itemize}

\begin{remark}[Accuracy vs Inaccuracy]
Dijkstra’s algorithm finds paths with the lowest cost, so cost function variables must reflect that principle: lower values indicate better outcomes. To address this, we use inaccuracy (the inverse of accuracy).
\end{remark}

The steps presented in 1)-4) illustrate how we transform the service graph to satisfy the requirements for applying Dijkstra's algorithm. We simplify this process by directly building the graph from the list of available services. Algorithm \ref{alg:create_graph}, outlines the procedure.

\begin{algorithm} 
\caption{CreateServiceGraph}
\label{alg:create_graph}
\begin{algorithmic}[1]
\Require services, cost function
\Ensure service graph

\State Initialize service graph
\State Add \textit{start} to \textit{service graph}
\State Add all \textit{sensors} from \textit{services} to \textit{service graph} \State Connect \textit{start} to \textit{sensors}
\For{each \textit{filter} in \textit{services}}
    \If{\textit{filter} requires a \textit{model}}
        \For{each \textit{model} in \textit{services}}
            \State Add \textit{filter} with \textit{model} to \textit{service graph}
            \State Connect \textit{sensors} to \textit{filter}
        \EndFor
    \Else
        \State Add \textit{filter} to \textit{service graph}
        \State Connect \textit{sensors} to \textit{filter}
    \EndIf
\EndFor
\For{each \textit{controller} in \textit{services}}
    \If{\textit{controller} requires a \textit{model}}
        \For{each \textit{model} in \textit{services}}
            \State Add \textit{controller} with \textit{model} to \textit{service graph}
            \For{each \textit{filter}}
                    \If{\textit{filter} is compatible with \textit{controller}}
                        \State Connect \textit{filter} to \textit{controller}
                    \EndIf
            \EndFor
        \EndFor
        \Else
            \State Add \textit{controller} to \textit{service graph}
        \State Connect \textit{filters} to \textit{controller}
    \EndIf
\EndFor
\State Add \textit{actuator} nodes from \textit{services} to \textit{service graph}
\State Connect \textit{controllers} to \textit{actuators}
\State Add \textit{target} to \textit{service graph} 
\State Connect \textit{actuators} to \textit{target}
\State Add weights to edges according to cost function
\State \textbf{Return} \textit{service graph}

\end{algorithmic}
\end{algorithm}
First, we initialize the service graph (line 1) and add the start node (line 2). Then we add every sensor node (line 3) and connect them with the start node (line 4). Next we add a node for every filter (line 5-15). If the filter requires a model, we add every possible combination of the filter with the available models as separate nodes. We connect all the filters to all the sensors. For the controllers we do the same (line 16-30). However, we only connect controller and filter nodes if their models are compatible, which means the model complexity of the filter is higher than or equal the model complexity of the controller. After that we add the actuator nodes (line 31) and connect them to the controller nodes (line 32). We add the target node (line 33) and connect it to every actuator node (line 34). Lastly, we compute the cost for choosing each service and add it as a weight to all its incoming edges (line 35).

\subsection{Graph-Based Orchestration}
\label{sect:orchworkflow}
\begin{algorithm}
\caption{Orchestrate}
\label{alg:orchestrator_workflow}
\begin{algorithmic}[1]
\Require services, cost function
\State \textit{service graph} $\gets$ CreateGraph(services, cost function)
\State \textit{optimalPath} $\gets$ Dijkstra(service graph, start, target)
\State set \textit{optimalPath} as new control system architecture

\end{algorithmic}
\end{algorithm}

Given the method for creating the service graph we can now outline the process of orchestrating the system in Algorithm \ref{alg:orchestrator_workflow}. The first step is the creation of the service graph, by conducting Algorithm \ref{alg:create_graph}. Once this graph is compiled, the orchestrator determines the optimal service composition by applying Dijkstra’s algorithm. The resulting path is then set as the control system architecture. This process repeats every time the cost function changes or a service is removed, updated, or added.

\section{Evaluation}
\label{sect:evaluation}
To evaluate the graph-based orchestrator, we use a three-tank system where we control the the water level in tank three, subjected to the maximum height of the tank as a constraint. This system is used to assess three evaluation scenarios (i) the creation of a service graph and discovery of the optimal control system architecture, (ii) the control system’s response to the addition of a previously unknown service, and (iii) the control system’s response to changes in its objective. We use the cost function defined in equation \eqref{eq:cost_function}.
\begin{figure*}
	\centering
		\includegraphics[width=0.80\textwidth]{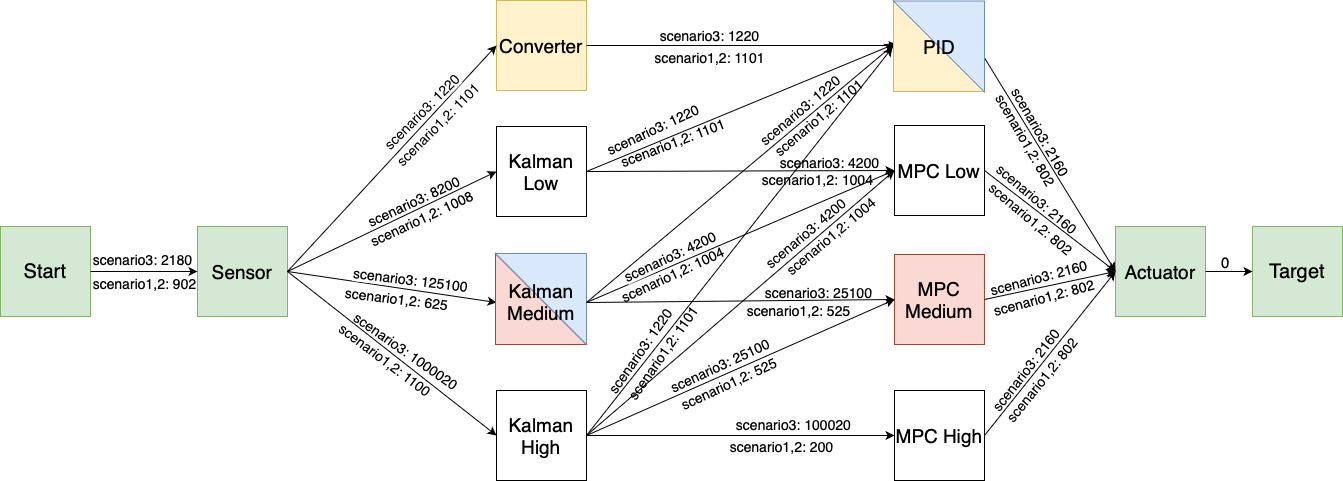}
	\caption{Control system architecture graph and the shortest path based on the values from Table \ref{tab:first}, the cost function from Eq.~\eqref{eq:cost_function} with $\alpha_\text{comp}=1$, $\beta_\text{comp}=100$ for scenarios 1, 2 and $\alpha_\text{comp}=1000$, $\beta_\text{comp}=20$ for scenario 3 using Dijkstra's algorithm where green nodes are common for all evaluation scenarios, blue for the first scenario, red for the second, yellow for the third scenario, and white for unselected services.}
	\label{fig:second_color3}
\end{figure*}
The \textit{computation factor} $x_\text{comp}$ is chosen based on the computation time of the services. The \textit{inaccuracy} $y_\text{inacc}$ is linked to the model used: lower inaccuracy implies a more complex model. To compute the \textit{inaccuracy} of a service's model, we make following assumptions:

\begin{itemize}
    \item If the service is a controller then: 
    \begin{align}
    x_{\text{comp}} &= i^2: i \text{ is the control model inaccuracy, } i>1
    \label{eq:ctrl}
    \end{align}
    \item If the service is a filter, then:
    \begin{align}
    x_{\text{comp}} &= j^3: j \text{ is the filter model inaccuracy, }j>1
    \label{eq:fltr}
    \end{align}
\end{itemize}

The first step of the evaluation is to create a service graph from the available services and discovering the optimal control system architecture. Table \ref{tab:first} lists the available services and their attributes relevant to the cost function. For simplicity, we assume one available sensor and actuator service, three models, four filters and four controllers. Three of the filters consist of a Kalman filter combined with models of different complexity. The remaining filter is a converter that simply forwards the values from the sensor. Similarly, three of the controller services are \gls{MPC} combined with models of different complexities and the remaining controller is a PID controller. 
\begin{table}[!htb]
    \centering
    \caption{Services and their attributes}
    \begin{tabular}{|c|c|c|}
        \hline
        \textbf{Service Type} & \textbf{$x_\text{comp}$} & \textbf{$y_\text{inacc}$} \\
        \hline
        Sensor & 2 & 9 \\ \hline
        Kalman Filter & see Eq.~\ref{eq:fltr} & low/medium/high model\\ \hline
        Converter Filter & 1 & 11 \\ \hline
        Model Low Complexity & 2 & 10 \\ \hline
        Model Medium Complexity & 5 & 5 \\ \hline
        Model High Complexity & 10 & 1 \\ \hline
        PID Controller & 1 & 11 \\ \hline
        \gls{MPC} & see Eq.~\ref{eq:ctrl} & low/medium/high model \\ \hline
        Actuator & 2 & 8 \\ \hline
    \end{tabular}
    \label{tab:first}
\end{table}
For the first scenario, we exclude the \gls{MPC} controllers. Thus, the orchestrator only has to choose the optimal filter model. For scenarios 1 and 2 we set the weighting coefficients to $\alpha_\text{comp}=1$ and $\beta_\text{comp}=100$.

The orchestrator executes Algorithm \ref{alg:orchestrator_workflow} to create the service graph and compute the optimal service configuration in terms of the cost function given in Equation \eqref{eq:cost_function}. Figure \ref{fig:second_color3} shows the generated weighted service graph and the shortest path for scenarios 1, 2, and 3. Green nodes represent start, sensor, actuator, and target, common to all three evaluation scenarios. Blue nodes indicate the selected filter and controller for scenario 1, where all \gls{MPC} nodes are excluded. We can observe that the orchestrator selects the Kalman filter with medium model complexity, which has the lowest cost of 625 out of all filters. The PID controller had to be used, since the \gls{MPC} was not active in this scenario.

In the second scenario we simulate the control system's response to adding a previously unknown service, namely the \gls{MPC}. The cost function remains unchanged from scenario 1. The orchestrator uses algorithm \ref{alg:orchestrator_workflow} to create the updated service graph and determines the optimal control system architecture. Figure \ref{fig:second_color3} shows the resulting weighted service graph and optimal control system architecture, with nodes colored in green and red. We see that the addition of the new controller has altered the optimal control system architecture, although the coefficients of the cost function have remained the same. While the Kalman Filter with medium model complexity prevails as the optimal choice, the chosen controller service moved from the PID in scenario 1 to the \gls{MPC} with medium model complexity in scenario 2 with a cost of 525. Furthermore, the \gls{MPC} with high model complexity has an even lower individual cost of 200, but is not compatible with the Kalman filter of medium model complexity. The combined cost of the Kalman filter and \gls{MPC} with high model complexity is higher than the combined cost of the Kalman Filter and \gls{MPC} with medium complexity.

Last, scenario 3 demonstrates the control system’s flexibility by altering the control system architecture in response to a change in the system’s objective. This adjustment is made by modifying the factors $\alpha_\text{comp}$ and $\beta_\text{comp}$ based on the importance of their attributes. Previously, attribute $y_\text{inacc}$ was prioritized over $x_\text{comp}$. For scenario 3 we set the weights in the cost function to $\alpha_\text{comp}=1000$ and $\beta_\text{inacc}=20$, notably increasing the importance of the computation time compared to the inaccuracy. Algorithm \ref{alg:orchestrator_workflow} is executed by the orchestrator, producing the weighted service graph, shown in Figure \ref{fig:second_color3}. The service graph is updated for the new cost function and the new optimal control system architecture is computed. The impact of the changed cost function is clear even with the same services. The updated control system architecture with green and yellow nodes is shown in Figure \ref{fig:second_color3}. Due to the much higher \textit{computation factor} the converter filter and the PID controller were chosen.

\section{Conclusion and Future Work}
\label{sect:conclusion}
This work introduces a graph-based orchestration method for model-based control system architecture using a service-oriented architecture. Our approach significantly enhances adaptability and flexibility by enabling the addition of new services at runtime, dynamic changes to the control system's architecture and identifying the optimal control system architecture in terms of a chosen cost function. The evaluation shows how this method allows for a greater variety in dynamic control system architectures and optimizes the system’s performance by selecting the best configuration based on the current conditions.

Future work includes extending the cost function by adding a latency variable for data transmission times between services. Additionally, automatic adjustment of the weighting coefficients based on the system state would improve performance. Further, placing greater emphasis on the inaccuracy factor when the system is near unsafe areas.

\section*{Acknowledgment}
The authors wish to express their sincere thanks to
Ole Greß and Thomas Stewens for the valuable discussions.

\end{document}